\documentclass[aps,prl,twocolumn,superscriptaddress,english,graphics]{revtex4}
\usepackage{graphicx,color}
\usepackage{amssymb,amsmath,amsfonts,mathrsfs,eufrak,fancyhdr}
\usepackage{hyperref}
\usepackage{pdfpages}
\usepackage{xcolor}

\begin{document}

\title{Quantum Jamming: Critical Properties of a Quantum Mechanical Perceptron}
\author{Claudia Artiaco}
\author{Federico Balducci}
\affiliation{SISSA, via Bonomea 265, 34136, Trieste, Italy}
\affiliation{The Abdus Salam International Center for Theoretical Physics, Strada Costiera 11, 34151, Trieste, Italy}
\affiliation{INFN Sezione di Trieste, Via Valerio 2, 34127 Trieste, Italy}
\author{Giorgio Parisi}
\affiliation{Dipartimento di Fisica, Sapienza Universit\`a di Roma, P.le A. Moro 2, 00185 Roma, Italy}
\affiliation{INFN, Sezione di Roma I, P.le A. Moro 2, 00185 Roma, Italy}
\affiliation{Nanotec-CNR, UOS Rome, Sapienza Universit\`a di Roma, P.le A. Moro 2, 00185, Roma, Italy}
\author{Antonello Scardicchio}
\affiliation{The Abdus Salam International Center for Theoretical Physics, Strada Costiera 11, 34151, Trieste, Italy}
\affiliation{INFN Sezione di Trieste, Via Valerio 2, 34127 Trieste, Italy}

\begin{abstract}
    In this Letter, we analyze the quantum dynamics of the perceptron model: a particle is constrained on a $N$-dimensional sphere, with $N\to \infty$, and subjected to a set of randomly placed hard-wall potentials. This model has several applications, ranging from learning protocols to the effective description of the dynamics of an ensemble of infinite-dimensional hard spheres in Euclidean space. We find that the jamming transition with quantum dynamics shows critical exponents different from the classical case. We also find that the quantum jamming transition, unlike the typical quantum critical points, is not confined to the zero-temperature axis, and the classical results are recovered only at $T=\infty$. Our findings have implications for the theory of glasses at ultra-low temperatures and for the study of quantum machine-learning algorithms.
\end{abstract}

\maketitle

\textit{Introduction.---} Constraint Satisfaction Problems \cite{mezard2009information} (CSPs), born in computer science, have taken up a prominent role also in statistical mechanics. Methods from the theory of disordered systems have been proposed to shed light on the possible origin of their computational difficulty \cite{kirkpatrick1994critical,monasson1999determining,mezard2001bethe,hartmann2005phase,krzakala2007gibbs} and inspire efficient algorithms \cite{mezard2002analytic} to solve them. While problems defined in terms of discrete variables map naturally to spin glasses, CSPs with continuous variables have shown a deep connection with structural glasses \cite{liu1998nonlinear,torquato2000random,charbonneau2012universal,charbonneau2014fractal,charbonneau2017glass,berthier2019gardner}. 

A notable example of CSP with continuous variables is the sphere packing problem \cite{franz2016simplest,franz2017universality}. Sphere systems have gained plenty of attention among the glass physics community, and their jamming transition has been incorporated into the framework of glassy theory \cite{charbonneau2014fractal}. In this context, the perceptron, another CSP with continuous variables, has been recognized as the simplest mean-field model presenting a jamming transition in the same universality class of high-dimensional sphere systems \cite{franz2015universal,franz2016simplest,altieri2016jamming,franz2017universality}. Exactly solvable models have always played an important role in increasing our understanding of the physics of glasses, both qualitatively and quantitatively. Furthermore, the perceptron has several applications in learning protocols \cite{gardner1988space,gardner1988optimal,battista2020capacity} and constitutes the building block of deep neural networks.

Recently, partly motivated by the technological progress in quantum computation \cite{nielsen2000quantum}, many authors have been looking at ways to use quantum dynamics to speed up the solution of the classical problems. In the case of discrete variable CSPs, a growing body of literature has investigated the impact of quantum dynamics on the spin-glass transition \cite{bray1980replica,miller1993zero,guo1994quantum,rieger1994zero,read1995landau,laumann2008cavity}, and it has been found that disordered quantum systems display a plethora of new phenomena, such as Many-Body Localization (MBL) \cite{altshuler2010anderson,laumann2014many,laumann2015quantum,nandkishore2015,PhysRevB.93.024202,mossi2017ergodic,abanin2017,imbrie2017,chandran2015constructing,baldwin2018quantum,smelyanskiy2020nonergodic}. The study of CSPs with continuous variables endowed with quantum dynamics, surprisingly, has not received the same kind of attention so far, but it promises to be equally far-reaching. For instance, in view of the connection to structural glasses, it might provide clues for the anomalous (i.e.\ non-Debye) behavior of thermodynamic quantities in glasses at ultra-low temperatures. These phenomena, such as $C_V(T\sim0)\sim T$ \cite{zeller1971thermal,jackle1972ultrasonic,perez2014two}, are indeed naturally explained in terms of quantum mechanics \cite{anderson1972anomalous,phillips1972tunneling}, but no firm results or solvable toy models exist (see for example \cite{leggett2013tunneling} for a critical view).

The purpose of this Letter is to address, for the first time, the jamming transition deep in the quantum regime \cite{nussinov2013mapping} through the perceptron model. We show that quantum mechanical effects change the nature of the critical phase radically. We find that, for any $\hbar \neq 0$, the critical exponents are different from the classical ones and independent of the temperature. We also find that $C_V(T \sim 0)\sim e^{-\Delta/T}$ at small $T$, while at higher temperatures the specific-heat has a power-law behavior. Remarkably, the latter result, valid in the deep quantum regime, resembles the semiclassical results of Refs.\ \cite{Franz13768,schehr2005low}, connecting the physics on the two sides of the jamming transition \footnote{We notice that the name {\it quantum perceptron} has been already used in the past for a quantum algorithm for learning quantum states \cite{lewenstein1994quantum}. Our problem is different since we implement a quantum dynamics on a classical constraint satisfaction problem.}.

\begin{figure}
    \includegraphics[scale=0.3]{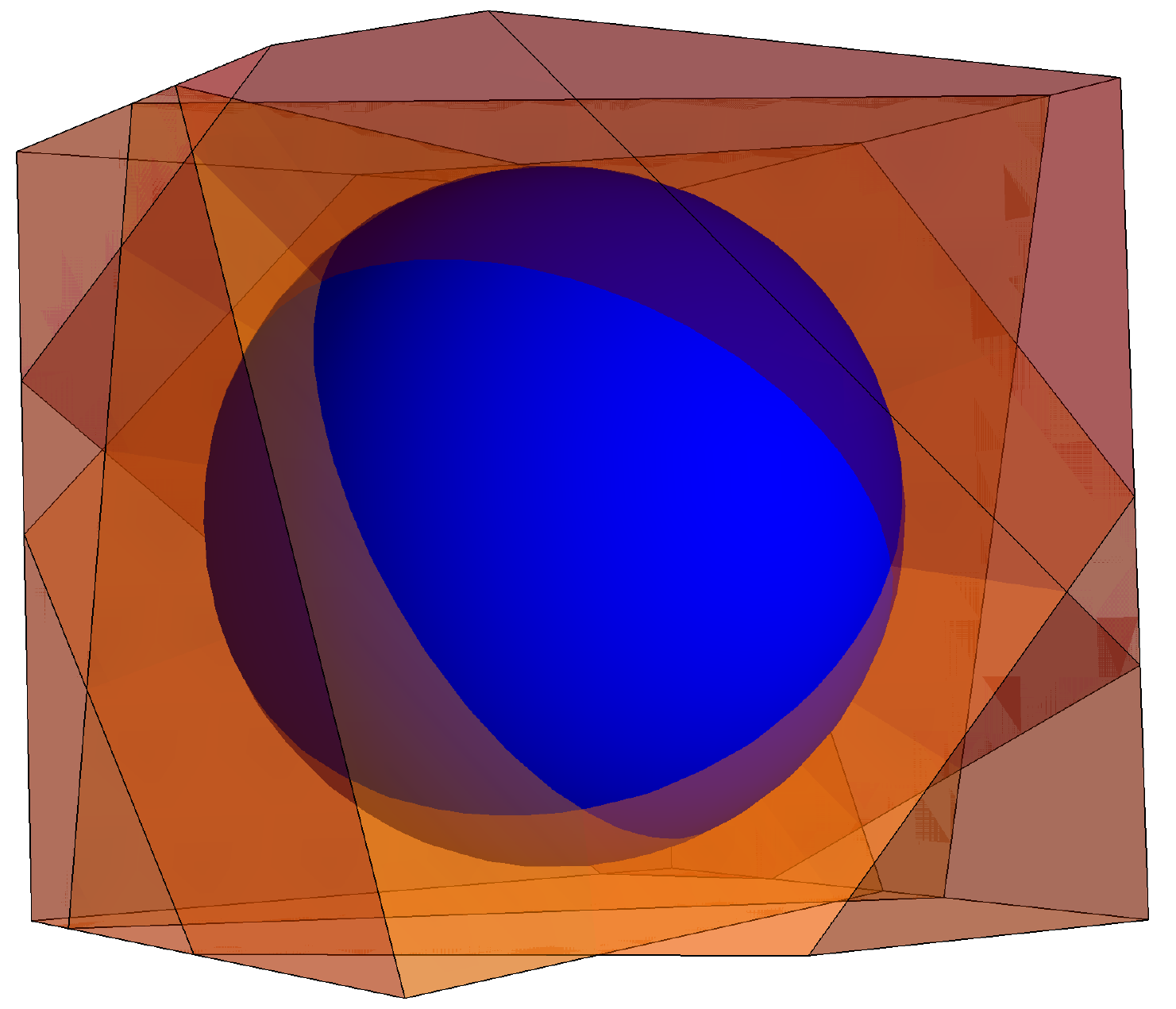}
    \caption{Finite dimensional representation of the perceptron model at $\sigma =0$, $N=3$, $M=4$. Each constraint is represented by a plane passing through the origin, and cuts the sphere in half. The particle can move in the allowed (light blue) region. The jamming transition is reached when the number of obstacles is such that (with probability 1 in the $N,M\to\infty$ limit) there is no light blue volume left anymore.}
    \label{fig:sphere}
\end{figure}

\textit{Model.---} The perceptron model can be formulated as a particle living on a $N$-dimensional sphere, subjected to a set of randomly placed obstacles. The vector $\boldsymbol{X}$ represents the position of the particle on the sphere ($\boldsymbol{X}^2=N$), and the obstacles are represented by the $N$-dimensional vectors $\boldsymbol{\xi}^{\mu} = (\xi^{\mu}_1,\dots,\xi^{\mu}_N)$, where $\mu = 1,\dots, M=\alpha N$ and $\xi^\mu_i$ are i.i.d. Gaussian random variables with zero mean and unit variance. For each obstacle, one defines the constraint
\begin{equation}
    h_{\mu}(\boldsymbol{X})=\frac{1}{\sqrt{N}}\boldsymbol{\xi}^{\mu}\cdot\boldsymbol{X}-\sigma > 0,
\end{equation}
and the cost function is $V=\sum_{\mu=1}^Mv(h_{\mu}(\boldsymbol{X}))$. We are interested in the hard-wall potential case in which $v(h)=0$ if $h>0$ and $v(h)=\infty$ if $h<0$; hence, all the constraints must be satisfied (see Fig.\ \ref{fig:sphere}). The limits $N,M\to\infty$ are taken, eventually, keeping $\alpha \equiv M/N$ finite.

The classical system (recovered for $\hbar= 0$) is independent of the temperature and presents two phases, determined by whether there is or there is not any volume left by the intersection of the $M$ constraints. More specifically, one has to consider the limit of the set $\mathcal{W} \equiv \bigcap_{\mu=1}^M \{\boldsymbol{X}\in \mathbb{R}^N: \boldsymbol{X}^2=N \wedge h_{\mu}(\boldsymbol{X})>0\}$ as $N,M \to \infty$: in the satisfiable (SAT) phase, a position $\boldsymbol{X}$ for the particle satisfying all the constraints can be found with probability one. In the unsatisfiable (UNSAT) phase, instead, $\mathcal{W}$ becomes empty and the CSP problem has no solution. The sharp SAT-UNSAT transition is induced by increasing the constraint density $\alpha$ up to $\alpha_c(\sigma)$.

The features of the SAT-UNSAT transition depend on $\sigma$ \cite{franz2017universality}. For $\sigma > 0$, the constraints $\{ h_\mu \!> \! 0\}$ force the particle $\boldsymbol{X}$ to be closer than some distance to each obstacle; thus, the allowed region is convex. The free energy has a single minimum and the replica-symmetric (RS) solution is everywhere stable. On the contrary, when $\sigma<0$, the constraints are satisfied if the particle is away from each obstacle. The allowed region is non-convex and can be composed of disconnected islands. The SAT-UNSAT transition falls within a phase where the landscape is rugged and marginally stable. For our purposes, we will concentrate only on the value $\sigma=0$, at the border of the RS stable region, for which the jamming point correponds to $\alpha_c(0)=2$. In this way, we can reach the jamming point within the RS ansatz, but capturing the physics of the glassy phase ($\sigma < 0$) .

The model is quantized by imposing the canonical commutation relations $[\hat{X}_i,\hat{P}_j]=i\hbar\delta_{ij}$. The Hamiltonian reads
\begin{equation}
    \label{eq:perc-hamiltonian}
    \hat{H}=\frac{\hat{\boldsymbol{P}}^2}{2m}+\sum_{\mu=1}^Mv(h_{\mu}(\hat{\boldsymbol{X}})).
\end{equation}

\textit{Methods.---}We wish to compute the quenched disorder average of the free energy $F=-\beta^{-1}\overline{\ln Z}$, $Z={\rm Tr}(e^{-\beta H})$. Following the lengthy but straightforward procedure in \cite{Franz13768,SupplMat}, which introduces replicas $a,b=1,...,p$ (with eventually $p\to 0$), the quenched free energy within the RS ansatz is expressed in terms of the Edwards-Anderson order parameter $q=N^{-1}\overline{\langle\boldsymbol{X}_a(t)\cdot\boldsymbol{X}_b(s)\rangle}$ (for $a\neq b$ and any $t,s$), the correlation function of a single replica $G(t-s)=N^{-1}\overline{\langle \boldsymbol{X}_a(t)\cdot\boldsymbol{X}_a(s)\rangle}-q$, and a Lagrange multiplier $\mu$ to enforce the spherical constraint. $G(t), q,$ and $\mu$ need to be found self-consistently. To this purpose, it is convenient to introduce a one-dimensional, $\beta \hbar$-periodic auxiliary process with the same autocorrelation function
\begin{equation}
    \langle\bullet\rangle_r= \frac{1}{Z_0} \oint Dr \, e^{-\frac{1}{2}\iint_0^{\beta\hbar}\frac{dt}{\beta\hbar}\frac{ds}{\beta\hbar}r(t)G^{-1}(t-s)r(s)}\bullet
\end{equation}
($Z_0$ is a suitable normalization). Then, the quenched free energy in the RS approximation, per dimension and per replica, reads \footnote{This quantity, as it is written, is divergent. In Ref.\ \cite{Franz13768} it is shown how to properly regularize it and renormalize it. In particular, the thermodynamic observables, like specific heat and order parameter, are divergence-free.}

\begin{gather}
\notag
    - \beta f =
    \frac{1}{2}\sum_{n\in\mathbb{Z}}\ln{G_n}+\frac{q}{2G_0}-\frac{\beta m}{2}\sum_{n\in\mathbb{Z}}\omega^2_nG_n\\
    -\frac{\beta \mu}{2}\Big[\sum_{n\in\mathbb{Z}}G_n-(1-q)\Big]\nonumber\\
    +\alpha\gamma_q \star\ln{\langle e^{-\beta\int_0^{\beta\hbar}\frac{dt}{\beta\hbar}v(r(t)+h)}\rangle_r},
    \label{eq:RSfree-energy}
\end{gather}
where we denoted the Fourier transform by
\begin{equation}
    \bullet(\omega) \equiv \int_0^{\beta\hbar} \frac{dt}{\beta\hbar} e^{-i\omega t} \bullet(t), \quad
    \bullet_n \equiv \bullet(\omega_n)
\end{equation}
$\omega_n \equiv 2 \pi n / \beta \hbar$ being the Matsubara frequencies, and $\gamma_q\star\bullet(h)\equiv\int_{-\infty}^{\infty}\frac{dh}{\sqrt{2\pi q}}e^{-h^2/2q}\bullet(h)$.  It is also convenient to define the self-energy
\begin{equation}
    \Sigma(\omega)=\beta^{-1}G^{-1}(\omega)-m\omega^2-\mu,
\end{equation}
and fix $\Sigma(0)=0$.

As said before, the extremization of (\ref{eq:RSfree-energy}) with respect to $G_n,q,\mu$ gives rise to a set of self-consistency equations (see \cite{SupplMat}). To solve them we have implemented an iterative method, together with a Montecarlo sampling for the calculation of $\langle\bullet\rangle_{r}$, (for an analog calculation in the SK model see \cite{andreanov2012long,young2017stability,biroli2001quantum,cugliandolo2001imaginary}).

\begin{figure}
    \includegraphics[width=\columnwidth]{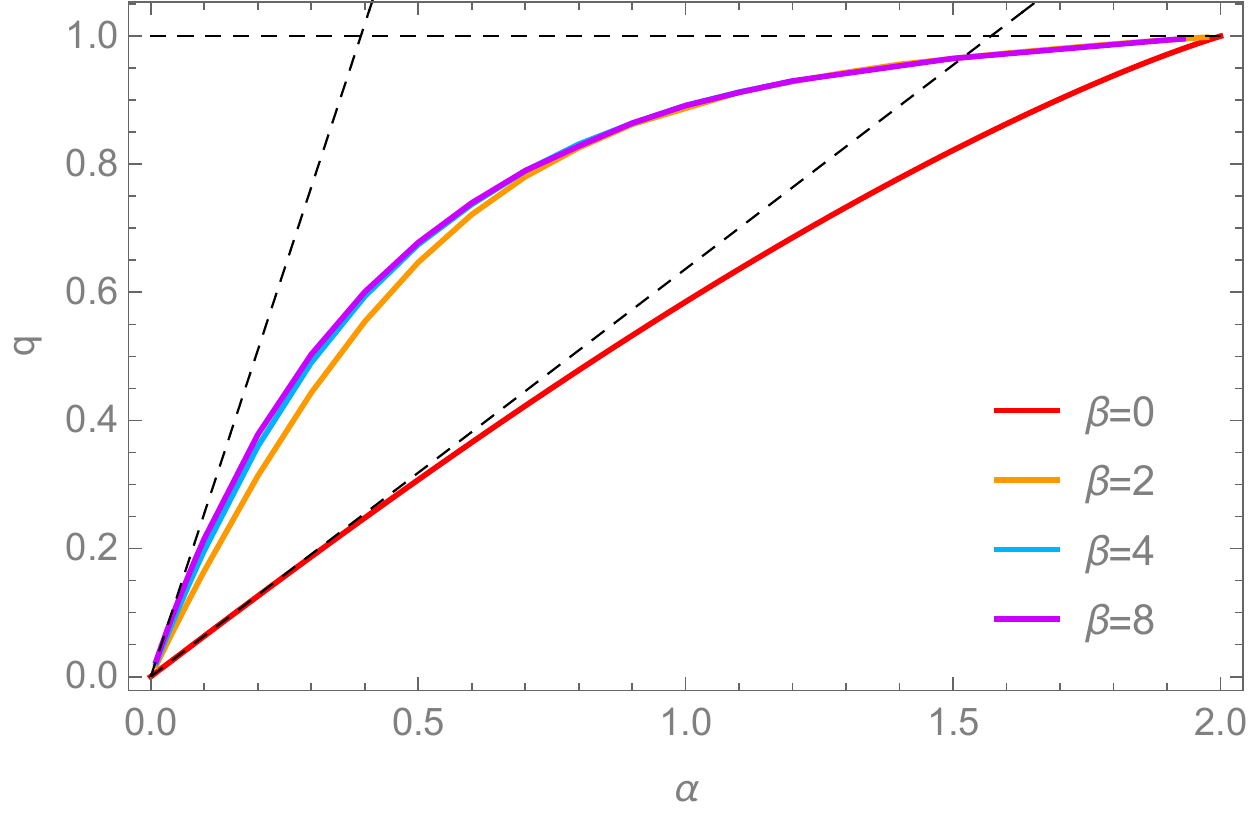}
    \caption{Edwards-Anderson order parameter as a function of the constraint density $\alpha$ for various temperatures. From bottom to top: infinite temperature classical dynamics (red) to finite temperature quantum dynamics ($\beta=2,4,8$). The $O(\alpha)$, $\beta=\infty$ results are shown as dashed black lines (while the horizontal black line is a reference for the value $q=1$). Notice how, as soon as $\alpha\gtrsim 1,$ the temperature dependence of $q$ is effectively lost (it is $\sim e^{-c\beta/(2-\alpha)^2}$).}
    \label{fig:QA}
\end{figure}

\begin{figure}
    \includegraphics[width=\columnwidth]{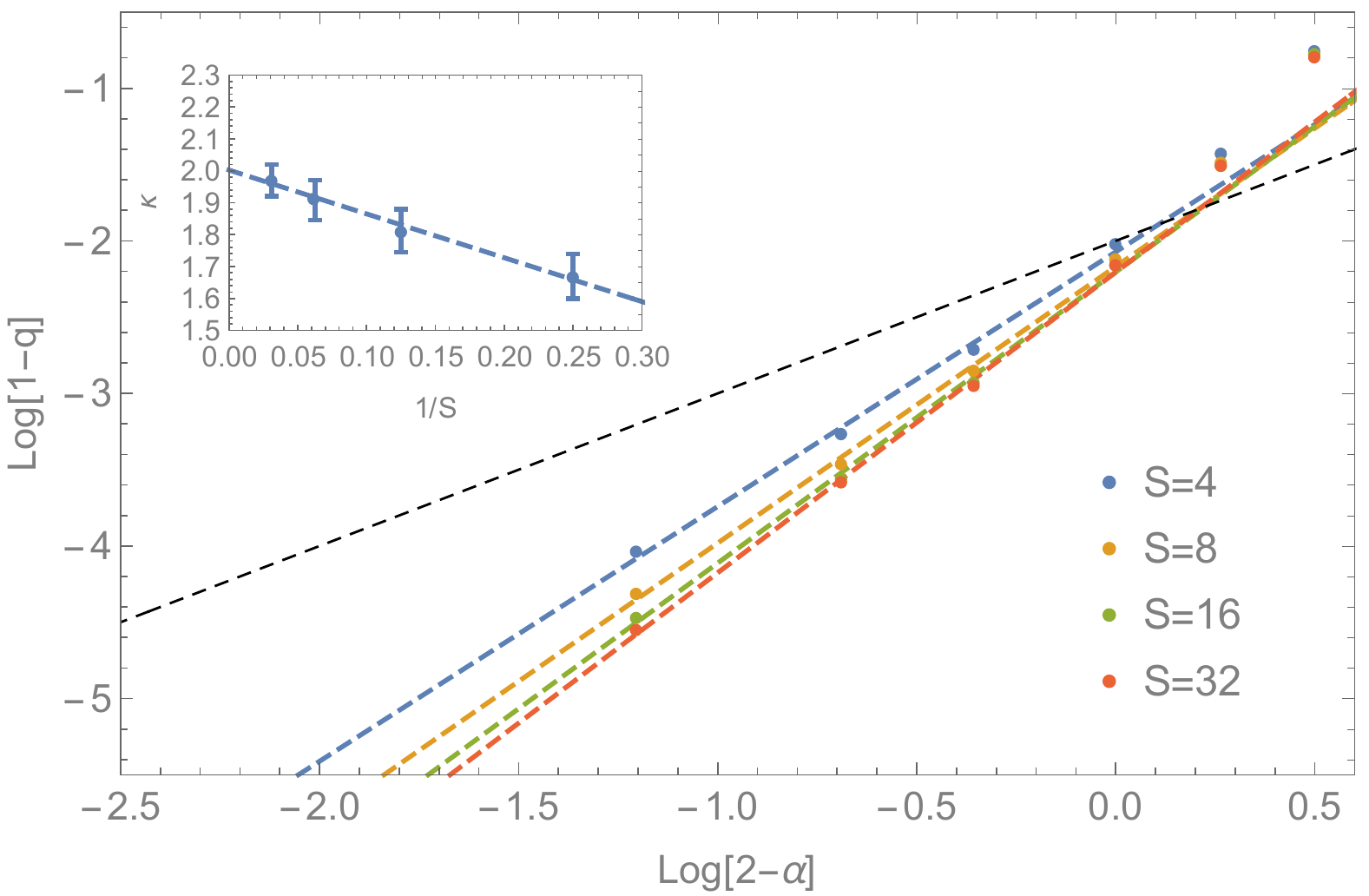}
    \caption{Edwards-Anderson order parameter close to the critical point $\alpha=2$. From top to bottom, increasing the number of Trotter slices $S=4,8,16,32$ for sufficiently large $\beta$, the slope increases. For reference, the classical value of the slope (from $(1-q)\sim (2-\alpha)$) is shown as the diagonal dashed black line. In the inset are shown the values of the slope with their errors, and its extrapolation to $S\to\infty$ to the value $\kappa=2.0\pm 0.1$, quoted in the text.}
    \label{fig:QAcrit}
\end{figure}

\textit{Results.---}As a first result, we obtain the value of the order parameter $q$ as a function of $\alpha,\beta$; it is plotted in Fig. \ref{fig:QA} against the classical counterpart $q_{\rm cl}(\alpha)$, obtained at $\hbar=0$ \cite{franz2017universality}. Unlike the quantum case, $q_{\rm cl}(\alpha)$ is independent of the temperature and goes to $1$, for $\alpha\to \alpha_c = 2$, with the critical exponent $\kappa_{\rm cl}=1$ (valid for $\sigma\geq 0$, while for $\sigma<0$ one has $\kappa_{\rm cl}=1.41574...$ \cite{franz2017universality}): $(1-q_{\rm cl}(\alpha))\simeq \frac{1}{4}(2-\alpha)$. The value of $q$ for $\hbar>0$ is always larger than the classical one, and this can be easily understood: the ground state of a quantum particle in a billiard is more concentrated than a flat distribution on the billiard table, because of the Dirichlet boundary conditions on the walls. Moreover, it becomes more concentrated the larger the aspect ratio of the billiard, namely if one of the sides is larger than the others. Quantitatively, one finds $q > q_{\rm cl}$ already at lowest order in $\alpha$. Indeed, from the self-consistency equations \cite{SupplMat}, $q = \alpha \langle r_0\rangle^2_{v(h=0)}+O(\alpha^2)$ where the average $\langle \bullet \rangle _{v(h=0)}$, when $\beta\to\infty$, indicates the expectation value over the ground state of a harmonic oscillator with a wall in the origin. This problem is easily solved and one finds $q= \frac{8}{\pi}\alpha+O(\alpha^2)$, to be compared with $q_{\rm cl}=\frac{2}{\pi}\alpha+O(\alpha^2)$.

Fig.~\ref{fig:QA} shows that the quantum order parameter $q$ depends on the temperature $T=1/\beta$ for $\alpha \lesssim 1$, and then, increasing $\alpha$, becomes independent of $T$ through a crossover. From the classical calculation, we expect that the typical linear size of the allowed region for the particle on the sphere vanishes as $\ell\sim\sqrt{1-q_{{\rm cl}}}\sim\sqrt{2-\alpha}$ for $\alpha \to 2$. Thus, as soon as the energy gap to the first excited state becomes larger than the temperature, i.e.\ roughly when $\frac{\hbar^2}{m (1-q_{\rm{cl}})} \sim \frac{\hbar^2}{m(2-\alpha)} \gtrsim T$, the quantum dynamics is {\it effectively at zero temperature} and the order parameter $q$ becomes independent of $T$. Moreover, in the following we will show that the gap, deep in the quantum regime, grows even faster than $(2- \alpha)^{-1}$ when $\alpha \to 2$. Since the quantum dynamics recovers the classical dynamics only when the de Broglie wavelength $\lambda_T \sim \hbar / \sqrt{mT} \ll \ell$, on approaching jamming quantum mechanics dominates. Hence, for any $T,\hbar, m$, as $\alpha\to 2$ one eventually enters a quantum critical regime, where quantum mechanics controls the dynamics and defines, among other things, novel critical exponents. The classical result is recovered only by taking the limit $T \to \infty$ before $\alpha\to2$.

The value of the critical exponent $\kappa$ regulating the relation $(1-q)\sim (2-\alpha)^\kappa$ in the quantum regime can be extracted by looking at the low-temperature, large-$\alpha$ data. As usual, a sufficiently large number of Trotter slices $S$ must be taken, and it increases as $\alpha \to 2$, making the numerical simulations more demanding. However, fortunately, the asymptotic region is reached already at $\alpha\gtrsim 1$. The data in Fig.\ \ref{fig:QAcrit} clearly show that the critical exponent of the quantum theory is not the classical one, $\kappa_{\rm cl}=1$, and it departs more and more from it as the number of Trotter slices is increased. We have performed a log-log fit to extract such critical exponent, in a region of $\alpha\in [1,1.7]$. Extrapolating $S \to \infty$, we find $\kappa=2.0\pm 0.1$ (Fig. \ref{fig:QAcrit}).

That $\kappa > 1$ in the quantum case can be understood also from a simple variational calculation \cite{SupplMat}. Using in the scaling region $\alpha \to 2$ the (uncontrolled) approximation $G_n^{-1} = \beta m(\omega_n^2+\hbar^2/4m^2)/(1-q)$, we were able to solve explicitly the self-consistency equations for $\beta \to \infty$, finding $\kappa=3/2$. The value $\kappa \simeq 2$ from the Montecarlo simulations presumably comes once the true behavior of $\Sigma(\omega)$ is considered.

The internal energy per degree of freedom $u$ (see \cite{Franz13768,SupplMat}) is independent of $\beta$, like $q$, already at $\alpha\gtrsim 1$. Extrapolating its behavior for infinite number of Trotter slices, the internal energy diverges as $u \sim \frac{\hbar^2}{m (2-\alpha)^2}$ for $\alpha\to 2$. This can be again interpreted in terms of reduced volume and uncertainty principle, and confirms the previous result $\kappa \simeq 2$.

\begin{figure}
    \includegraphics[width=\columnwidth]{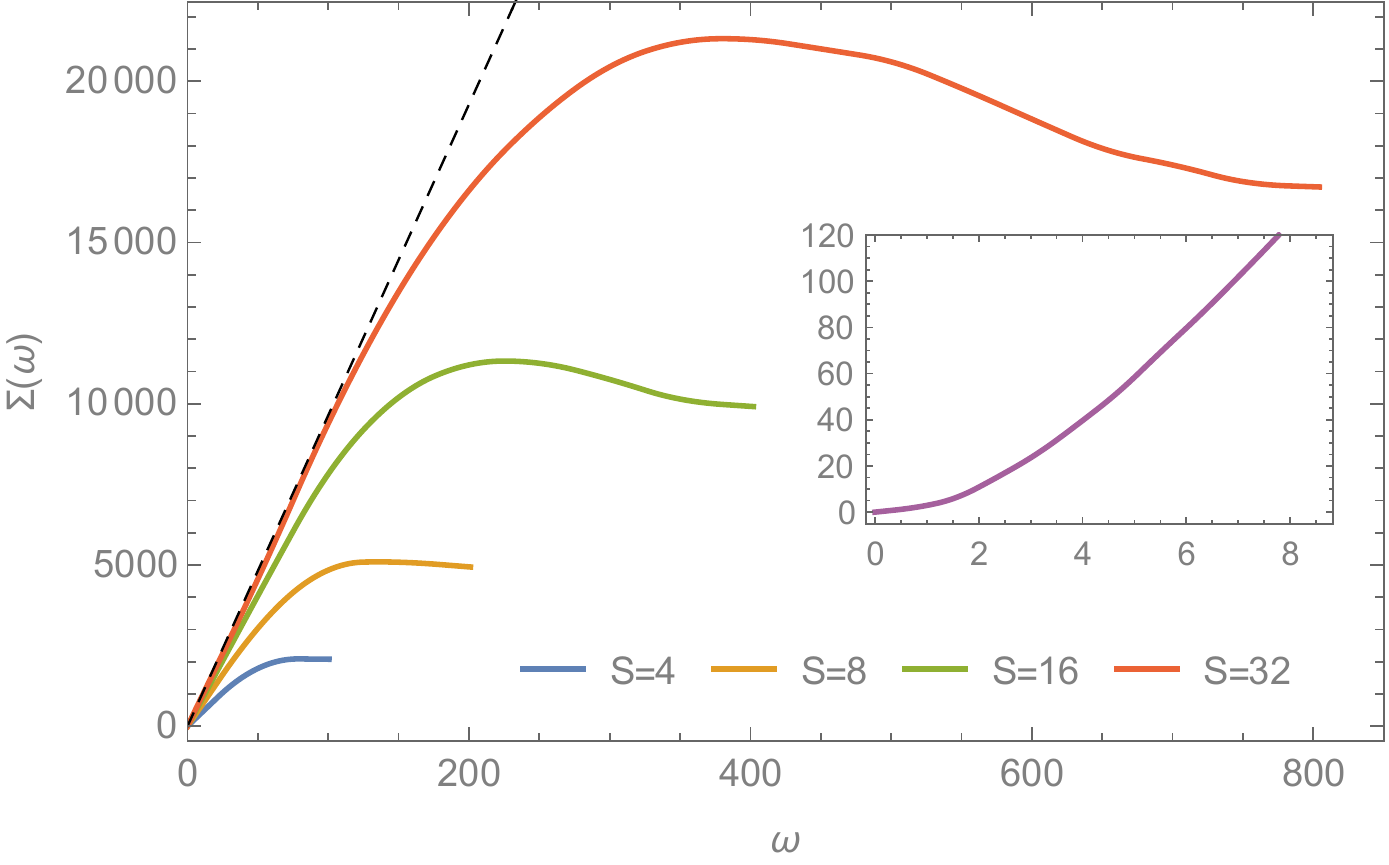}
    \caption{Self-energy $\Sigma(\omega)$ at $\alpha=1.7$, $\beta=1/2^3$ as a function of the Matsubara frequency $\omega$, for increasing number of Trotter slices (accessing higher and higher frequencies). We see that $\Sigma(\omega)$ develops a linear $\omega$ behavior (black, dashed line) for intermediate $\omega$'s, while retaining its analyticity in terms of $\omega^2$ around the origin for any $q<1$ (inset). In the inset, it is shown  $\Sigma(\omega)$ at small $\omega$'s for $\alpha=1.5$, $\beta=8$.}
    \label{fig:SelfEn}
\end{figure}

We have just shown that, at fixed temperature, in the quantum regime the critical properties of the system are determined by the ground state, and the gap to the first excited state grows as $\Delta \sim \frac{\hbar^2}{m (1-q)}$ for $\alpha\to 2$. This implies that, if we focus on frequencies $\omega\ll \Delta / \hbar$, or times $t\gg \hbar/\Delta$, there is no dynamics. In order to see some dynamical behavior one should consider $G(\omega\gtrsim \Delta/\hbar)$. As shown in Fig.\ \ref{fig:SelfEn}, at these large frequencies the form of the self-energy $\Sigma(\omega)$ changes significantly. Indeed, at any $\alpha<2$, the self-energy is an analytic function of $\omega^2$ in a neighborhood of the origin $\omega=0$ (inset of Fig.~\ref{fig:SelfEn}). As $\alpha\to 2$, this behavior becomes extended to increasing values of $\omega$. At larger frequencies, however, $\Sigma(\omega)$ develops a linear behavior. Moreover, for any $\alpha<2$, $\lim_{\omega \to \infty} \Sigma(\omega)=0$, as can be seen from its definition \cite{SupplMat}. Performing a log-log fit, we find that the constant contribution to the autocorrelation function scales as $\beta\mu\sim(1-q)^{\delta}$ where $\delta\simeq-0.9$. From a quadratic fit of $\Sigma(\omega)$ at small $\omega$, the coefficient of the quadratic term results instead almost independent of $(1-q)$.

The behavior of $\Sigma(\omega)$ defines the effective dynamics of the theory, and its analytical properties around the origin determine the low-temperature behavior of thermodynamical observables. Both the analyticity of $\Sigma(\omega)$ around $\omega=0$ and the independence of $\beta$ of all the observables, including the internal energy $u$, show that the specific heat is non-analytic in $T$ when $\alpha\to 2$. More precisely, our findings show that $C_V(T\sim 0) \sim e^{-\Delta/T}$, due to the presence of the gap. However, since at not-so-small $\omega$ it holds $\Sigma(\omega) \sim |\omega|$, the specific heat presents a power-law behavior at high enough temperatures, i.e. $C_V(T > T_{\rm cutoff}) \sim T^\gamma$. Since $\Delta \to \infty$ as $\alpha \to 2$, $T_{\rm cutoff} \to \infty$ too.

The linear dispersion $\Sigma(\omega)\sim |\omega|$, observed in the critical regime, reminds us of the result of \cite{Franz13768}, where the authors perform a semiclassical analysis to investigate the UNSAT phase with soft potentials. In \cite{Franz13768}, they sent $\hbar \to 0$ with $\hbar / T$ kept fixed, while in our study $\hbar$ is kept finite. They found the linear dispersion $\Sigma(\omega)\sim |\omega|$ in a neighborhood of the origin $\omega=0$, implying a power-law behavior of $C_V(T)$ at small $T$ near the jamming point. The similarities between the two results are surprising, since the regimes considered are different, and suggest that the linear dispersion $\Sigma(\omega)\sim |\omega|$ might be a universal feature of quantum models near jamming. 

\vspace{2mm}

\textit{Conclusions.---}We have investigated the quantum perceptron with hard-wall potentials as a model for jamming. We have studied the replicated, quenched free energy in the RS approximation, finding a quantum critical point corresponding to the classical jamming point $\alpha_c=2$ at $\sigma=0$. Usually, quantum critical points are confined and influence the physics around $T=0$ \cite{sachdev1999quantum}. We instead find that the quantum jamming critical point exists \emph{for any temperature}, and the classical results are recovered only by taking $T\to\infty$ before $\alpha\to\alpha_c$: it is the classical critical point to be confined to $T=\infty$. We find quantum critical  exponents different from the classical ones, and an exponentially small $C_V(T)$ at small $T$. The dispersion relation $G(\omega)^{-1}\sim |\omega|$ for frequencies higher than the gap, but not asymptotically large, implies a power-law specific-heat for $T>T_{\rm cutoff}$, where $T_{\rm cutoff}$ diverges at the critical point. This shows a surprising connection of our findings with the ones of the semiclassical analysis in \cite{Franz13768}, where a different region of parameters was considered, that deserves to be further investigated. 

An appealing extension of this work would be to consider soft potentials, having a finite $v'\equiv\partial v/\partial r|_{r= 0}$ \footnote{A similar reasoning applies for the case $v'=0$ but $\partial^k v / \partial r^k |_{r=0} \neq 0$ with $k>1$.},  as in the case of structural glasses. Employing soft potentials, it is possible to access the UNSAT phase deep in the quantum regime. We do expect that the quantum jamming transition will turn into a crossover (like the classical one does) but the same phenomenology outlined in this paper should be observed as far as the change in the potential on length scales $O((1-q)^{1/2})$ is large with respect to the gap $\Delta \sim (1-q)^{-1}$. This means that for $(1-q) \gtrsim (v')^{-2/3}$, or $\alpha \lesssim 2-c (v')^{-1/3}$, the physics is dominated by the hard-wall quantum jamming critical point. The robustness with temperature of the quantum critical point, shown in our results, implies that the quantum character of the system even with soft potentials cannot be neglected. Therefore, it suggests that the standard approaches used to study glassy systems at ultra-low temperatures, which add quantum effects on top of the classical landscape \cite{reinisch2004local,reinisch2005moving,khomenko2020depletion}, might be inadequate.

Another interesting extension of this study would be to move to the regions with $\sigma\neq0$. The case $\sigma>0$ is studied in learning protocols. Here, the same methods adopted in our study can be implemented, and one can directly investigate the effects of quantum dynamics. In the region $\sigma<0$, instead, it is also necessary to solve the self-consistency equations in the replica symmetry breaking framework. As the allowed volume becomes clustered, quantum effects may play a double role: for low disorder, tunneling may help the particle to explore many disconnected flat regions, and speed up the search of solutions (as it happens in the quantum random energy model \cite{mossi2017ergodic,baldwin2018quantum,smelyanskiy2020nonergodic}); for high disorder, Anderson Localization may take place, breaking ergodicity and changing significantly the classical phase diagram. The interplay of these behaviors, hard to be guessed, deserves a complete investigation.

We are very grateful to R. Fazio, S. Franz, T. Maimbourg, and D. Tisi for useful discussions and comments on a previous version of this paper. This work was supported by the Trieste Institute for the Theory of Quantum Technologies. This project has received funding from the European Research Council
(ERC) under the European Union’s Horizon 2020 research and innovation programme (Grant No. 694925, G.P.).

\bibliography{SG-bib}


\onecolumngrid
\section*{Supplementary Material}

\renewcommand{\theequation}{S\arabic{equation}}

\subsection{Derivation of the self-consistency equations}

We need to compute the quenched disorder average of the free energy $F=-\beta^{-1}\overline{\ln {\rm Tr}(e^{-\beta H})}$, with $H$ given by Eq.~(2) (main text). Introducing the imaginary time $t$, the Lagrange multiplier $\lambda$ associated to the constraint $\boldsymbol{X}^2=N$ and $p$ replicas, one can find $F$ as a function of the overlap matrix
\begin{equation}
    Q_{ab}(t,s)=\overline{\langle\boldsymbol{X}_a(t)\cdot\boldsymbol{X}_b(s)\rangle}/N,
\end{equation}
where $Q_{ab}(t,s)$ periodic in $t$ and $s$ with period $\beta\hbar$ and $a,b=1,...,p$ are replica indices. The quenched free energy $f$, per dimension $N$ and per replica $p$ can be written as:
\begin{equation}
    - \beta p f=\frac{1}{2}\ln{\det{\hat{Q}(t,s)}}+\frac{m}{2\hbar}\sum_a\int_0^{\beta\hbar}dt\,\partial^2_sQ_{aa}(t,s)|_{s=t}-\frac{m}{2\hbar}\sum_a\int_0^{\beta\hbar}dt\,\lambda_a(t)(Q_{aa}(t,t)-1)+\alpha\ln{\zeta},
    \label{eq-exponent-free-energy-replica-trick1}
\end{equation}
where
\begin{equation}
    \label{eq-zeta-free-energy-replica-trick}
    \zeta=\exp{\bigg(\frac{1}{2}\sum_{a,b}\iint_0^{\beta\hbar}\frac{dt}{\beta\hbar}\frac{ds}{\beta\hbar}Q_{ab}(t,s)\frac{\delta^2}{\delta r_a(t)\delta r_b(s)}\bigg)} \cdot \exp \bigg(-\frac{1}{\hbar}\sum_c\int_0^{\beta\hbar}dt\,v(r_c(t))\bigg) \bigg|_{r_c(t)=0}.
\end{equation}

The RS ansatz for the saddle point is:
\begin{equation}
    Q_{ab}(t,s)\overset{  \text{RS}}{=}[q_d(t-s)-q]\delta_{ab}+q
\end{equation}
where $q_d(t)-q$ is the autocorrelation function of a replica, while the off-diagonal order parameter $q$ is the analog of the Edwards-Anderson order parameter: It is the overlap of two different replicas. As usual, one shall send $p\to 0$ after computing the quantities involving $Q$.

We need to find the saddle point with respect to variations of $Q$, namely of $q_d(t)$ and $q$, and $\mu \equiv m \lambda$. To do this, it is convenient to define 
\begin{equation}
    G(t-s)\equiv q_d(t-s)-q.
    \label{eq-def-propagator}
\end{equation}
From the $\beta\hbar$-periodicity in imaginary time, we can consider as variables the countable set of Fourier components of $G(t)$, i.e.\ $\{G_n\}_{n \in \mathbb{Z}}$. Then, the quenched free energy in the RS approximation, per dimension $N$ and replica $p$, is (Eq.~(4), main text)
\begin{equation}
    - \beta f =
    \frac{1}{2}\sum_{n\in\mathbb{Z}}\ln{G_n}+\frac{q}{2G_0}-\frac{\beta m}{2}\sum_{n\in\mathbb{Z}}\omega^2_nG_n -\frac{\beta \mu}{2}\Big[\sum_{n\in\mathbb{Z}}G_n-(1-q)\Big]+\alpha\gamma_q \star\ln{\langle e^{-\beta\int_0^{\beta\hbar}\frac{dt}{\beta\hbar}v(r(t)+h)}\rangle_r},
\end{equation}
where $\omega_n \equiv 2 \pi n / \beta \hbar$ are the Matsubara frequencies, $\gamma_q\star\bullet(h)\equiv\int_{-\infty}^{\infty}\frac{dh}{\sqrt{2\pi q}}e^{-h^2/2q}\bullet(h)$ and
\begin{equation}
    \langle\bullet\rangle_r= \frac{1}{Z_0} \oint Dr \, e^{-\frac{1}{2}\iint_0^{\beta\hbar}\frac{dt}{\beta\hbar}\frac{ds}{\beta\hbar}r(t)G^{-1}(t-s)r(s)}\bullet.
\end{equation}

The saddle-point equations for the parameters $G_n$, $\mu$ and $q$ are 
\begin{gather}
    G^{-1}_n=\beta m\omega_{n}^2+\beta\mu+\beta\Sigma_{n}
    \label{eq:Gn-self-cons}\\
    \sum_{n}G_n =1-q
    \label{eq:G1q}\\
    q=\alpha\gamma_q\star\langle r_0\rangle^2_v
    \label{eq:q_eq}
\end{gather}
where we defined the self-energy
\begin{equation}
    \label{eq:self-energy}
    \Sigma_n \equiv \alpha\big(G^{-1}_n-G^{-2}_n\gamma_q\star(\langle r^*_nr_n\rangle_v-\delta_{n0}\langle r_0\rangle_v^2)\big) / \beta,
\end{equation}
with
\begin{equation}
    \langle\bullet\rangle_v= \frac{ \langle e^{-\beta\int_0^{\beta\hbar}\frac{dt}{\beta\hbar}v(r(t)+h)}\, \bullet \rangle_r}{\langle e^{-\beta\int_0^{\beta\hbar}\frac{dt}{\beta\hbar}v(r(t)+h)} \rangle_r} .
\end{equation}
These equations define self-consistently the dynamics of the auxiliary random process $r(t)$.

\subsection{Iterative solution of the self-consistency equations}

The self-consistency equations (\ref{eq:Gn-self-cons})-(\ref{eq:q_eq}) can be solved with an iterative method, together with a Montecarlo sampling. The parameters $G_n$, $\mu$ and $q$ can be initialized with arbitrary values. However, numerically it is convenient to proceed in a stepwise manner, from smaller to higher $\alpha$'s. The algorithm is composed by three steps. 

\begin{figure}[htb]
    \includegraphics[scale=0.28]{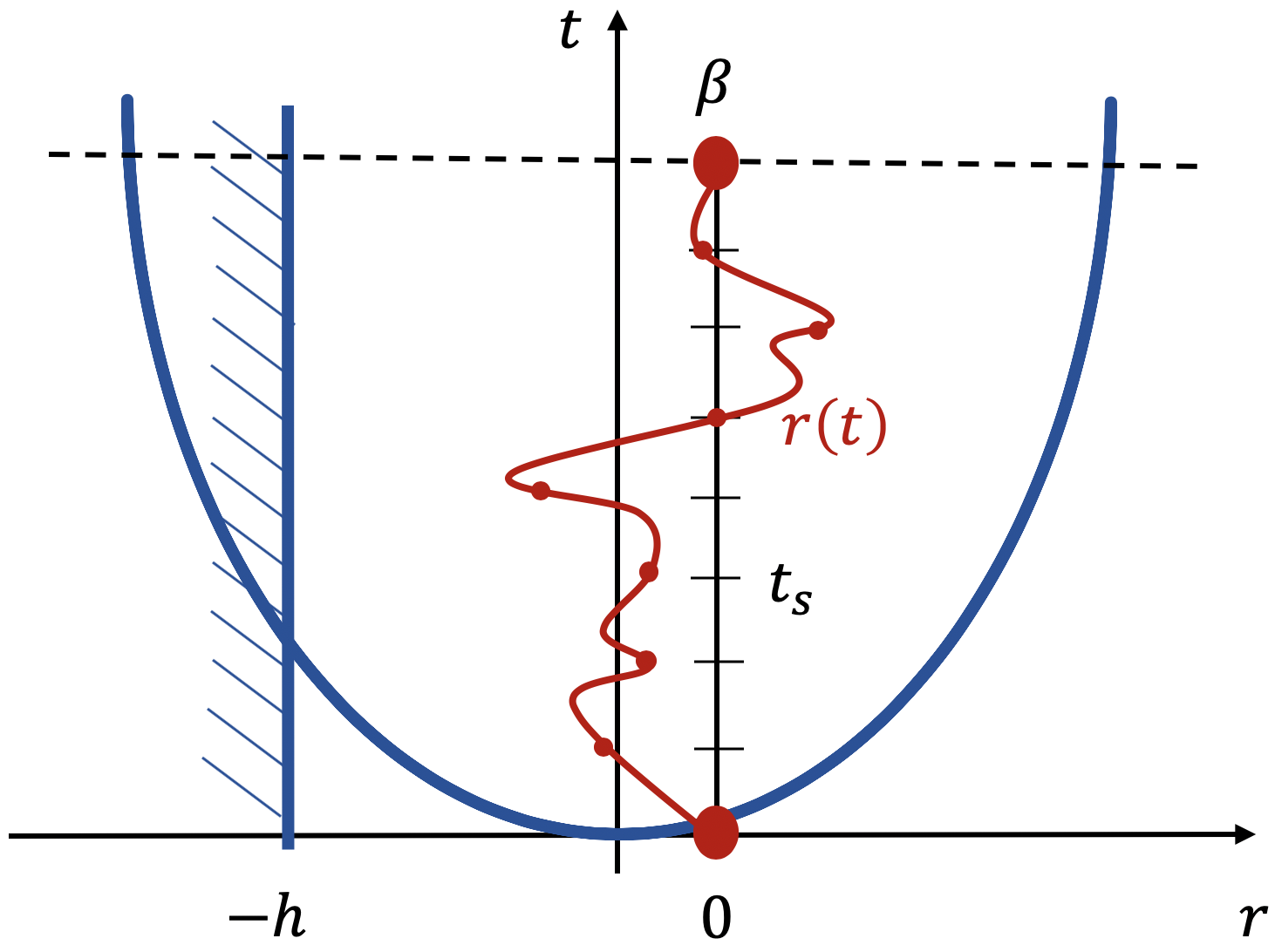}
    \caption{Sketch of the Path Integral Montecarlo (PIMC) used to simulate the dynamics of the  $\beta \hbar$-periodic process $r(t)$ ($r(0)=r(\beta\hbar)$) in the potential generated by $G^{-1}(t-s)$ and $v(r(t)+h)$. The PIMC consists in proposing a move $r(t_s) \to r(t_s) + \delta(t_s)$ for every time step $t_s$, which is accepted or rejected according to the Metropolis algorithm. We improved this simple scheme using both the method of images and the rigid movement of the time chain $r(t_s)$, as discussed in the text.}
    \label{fig:PIMC}
\end{figure}

The first step is to compute the self-energy $\Sigma_n$ and use it to update the autocorrelation function $G_n$, iteratively. However, the computation of the self-energy involves the averages $\langle\bullet\rangle_v$. To evaluate them, we use a Path Integral Montecarlo (PIMC) simulating the dynamics of the $\beta \hbar$-periodic process $r(t)$, in the potential generated by $G^{-1}(t-s)$ and $v(r(t)+h)$, as sketched in Fig.~\ref{fig:PIMC}. The former, when $\Sigma_n\equiv0$, contains a kinetic term ($m\omega_n^2/2$) plus a harmonic potential ($\mu/2$); the latter is the hard-wall potential which forces $r(t)>-h$. When $\Sigma_n \neq 0$ both contributions (kinetic and potential) change, and the dynamics of $r(t)$ becomes non-trivial. 

Numerically, it is convenient to consider the period of the process as $\beta$, reabsorbing $\hbar$ in the mass $m \to m/\hbar$. Moreover, the period has to be discretized: The number of Trotter slices is $S = \beta/a$, where $a$ is the time-slice amplitude, and, setting $\beta=2^L$ and $a=2^{-K}$, it holds $S=2^{L+K}$. In this way, we can define a discrete Fourier transform $f_n = \frac{1}{S} \sum_{s=0}^{S-1} f(t_s)e^{i \omega_n t_s}$ where $\omega_n = 2 \pi n/ \beta$ with $n\in[0,S-1]$. Thus, increasing $\beta$ the set $\{\omega_n\}$ becomes denser, while decreasing $a$ one can access higher frequencies.

The PIMC algorithm consists in proposing a move $r(t_s) \to r(t_s) + \delta(t_s)$ for every time step $t_s$, which is accepted or rejected according to the Metropolis algorithm with weight given by $G^{-1}$ and $v$. However, the presence of the hard-wall potential makes the convergence of the Montecarlo very demanding, and it is not sufficient to reject the attempted moves with $r<-h$ to have a good numerical protocol. Thus, we implemented an improved Montecarlo sampling which exploits the method of images. We modified the free particle kinetic term of the Hamiltonian ($m\omega_n^2/2$), and, instead of sampling the probability $P(r_0,0 \mid r_0, \beta)$ of the free particle, we used $P(r_0,0 \mid r_0, \beta)-P(\text{Im}(r_0),0 \mid r_0, \beta)$ where $\text{Im}(r_0)=-r_0-2h$ is the image of $r_0$ when the wall is in $-h$. Another expedient we adopted is to add a move which translates rigidly the time chain $r(t_s)$, i.e.\ $r(t_s) \to r(t_s) + \delta$ with $\delta$ independent of $t_s$.

The presence of the convolution $\gamma_q\star\bullet(h)$ in the definition of $\Sigma_n$ (Eq.\ \eqref{eq:self-energy}) implies the evaluation of $\langle \bullet \rangle_v$ for many positions $-h$ of the wall. We approximate this Gaussian integral with the Gauss-Hermite quadrature, always with, at least, 10 sample points. This first step of the iterative method stops when $G_n$ is converged for every $n$ within a fixed tolerance (we fixed the relative difference between $G_n^{\text{old}}$ and $G_n^{\text{new}}$ to be $<0.1\%$). 

The second step is to check if the converged $G_n$ verifies the identity in Eq.\ \eqref{eq:G1q}. If it does so, we can go to the third step; otherwise, $\mu$ is changed via the bisection method and the first step is performed again. 

The third step consists in computing the r.h.s. of Eq.\ \eqref{eq:q_eq} with the converged $G_n$ and $\mu$ and check if the identity in Eq.\ \eqref{eq:q_eq} is verified. If it is so, we have found the parameters which solve the self-consistency equations; if not, $q$ is changed and one has to repeat all the procedure from the first step.

\subsection{Exponent $\kappa=3/2$ in the quadratic approximation}

Setting $G_n^{-1} = \beta m (\omega_n^2+\hbar^2/4m^2)/(1-q)$ as in the text, the spherical constraint (Eq.\ \eqref{eq:G1q}) is automatically satisfied up to exponentially small corrections, and the values of $m$ and $q$ can be fixed by Eqs.~(\ref{eq:Gn-self-cons}) and (\ref{eq:q_eq}). Note that there is an equation of the form (\ref{eq:Gn-self-cons}) for every $n \in \mathbb{Z}$, yielding a deeply overcomplete set of constraints for our ansatz,  but we restrict to the $n=0$ case only.

It is convenient to set $x \equiv r/\sqrt{1-q}$ and $H \equiv h /\sqrt{1-q} $, so that Eq.\ \eqref{eq:q_eq} becomes
\begin{equation}
    \label{eq:q_eq_quadratic}
    \frac{q}{(1-q)^{3/2}} = \alpha \int \frac{dH}{\sqrt{2 \pi q}} \, e^{-\frac{(1-q)H^2}{ 2q}}\,
    \langle \psi_0^{(H)} | x | \psi_0^{(H)} \rangle^2 \,,
\end{equation}
where the reduced Schr\"odinger problem to solve is
\begin{equation}
    \label{eq:reduced-Schr}
    -\frac{1}{2}\frac{d\psi_k^{(H)}}{dx^2} + \frac{1}{8} x^2 \psi_k^{(H)} = E_k^{(H)} \psi_k^{(H)}, \;\; \psi_k^{(H)}(H) = 0.
\end{equation}
Self-consistently we will show that only the ground-state contribution matters (i.e.\ $k=0$). With this in mind we have employed the one-parameter variational wavefunction
\begin{equation}
    \psi^{(H)}(x;L) =\frac{1}{\sqrt{Z}}(x-H)\theta(x-H) e^{-x^2 / 4 L^2},
\end{equation}
with an appropriate normalization $Z$, for which the energy reads
\begin{equation}
    E^{(H)}(L) = \frac{1+L^4}{8L^2} \frac{\phi(H/\sqrt 2 L) (H^2 + 3L^2) - 2HL}{\phi(H/\sqrt 2 L) (H^2 + L^2) - 2HL}
\end{equation}
where $\phi(y) \equiv \sqrt{2 \pi} e^{y^2} {\rm Erfc}(y),$ ${\rm Erfc}$ being the complementary error function. The equation $d E^{(H)}/dL = 0$ can be solved separately in the regions $H \gg L$, $|H/L| \ll 1$ and $H \ll L$ by using suitable expansions. Remembering that $q \to 1$ (and therefore that the range of $H\sim \sqrt{q/(1-q)}\to \infty$) we see that the important region is $H\gg L$, and self-consistently we obtain $H/L\gg 1$. We find $\langle \psi_0^{(H)} | x | \psi_0^{(H)} \rangle\simeq H+3^{2/3}H^{-1/3}+O(H^{-5/3})$ and by inserting it in Eq.\ \eqref{eq:q_eq_quadratic} we arrive at
\begin{equation}
    \label{eq:q_eq2}
    q = \alpha \left[ (1-q) \xi\left(\frac{q}{1-q}\right) + \frac{q}{2} \right]
\end{equation}
with 
\begin{equation}
    \xi(\lambda) = \int_0^{\infty} \frac{d H}{\sqrt{2\pi \lambda}} \, e^{-H^2/2 \lambda} 
    \left[ \frac{(6H)^{2/3}}{2^{1/3}} + \cdots \right] = \frac{3^{2/3}\Gamma(5/6)}{\sqrt{\pi} 2^{1/3}} \lambda^{1/3} + \cdots\ .
\end{equation}
Eq.\ \eqref{eq:q_eq2} can now be solved for $q$, yielding $\kappa = 3/2$:
\begin{equation}
    q = 1 - \frac{\sqrt{2} \, \pi^{3/4} (2-\alpha)^{3/2}}{24 \, \Gamma(5/6)^{3/2}} \,.
\end{equation}
The same scaling has been observed by solving the Schr\"odinger equation \eqref{eq:reduced-Schr} numerically, discretizing the $x$-axis and employing imaginary-time evolution to find the ground state.

Knowing $q$ as a function of $\alpha$, we can now solve the $n=0$ case of Eq.\ \eqref{eq:Gn-self-cons} with the same technique. It reads
\begin{equation}
    m = \beta \gamma_{q/(1-q)} \star 
    \langle \psi_0^{(H)} | x^2 | \psi_0^{(H)} \rangle^{\rm conn} \, .
\end{equation}
By means of the same variational ansatz we find that the connected average in the equation above is $3^{1/3} H^{-2/3} \theta(H) + \cdots$, and finally 
\begin{equation}
    m = \beta \frac{3^{1/3}\Gamma(1/6)}{2^{5/6}} \left(\frac{1-q}{q}\right)^{1/3} \,.
\end{equation}
Thus we see that, as $q \to 1$, $\beta / m \to \infty$ and our approximation to take only the ground state becomes more and more reliable.

\subsection{Internal energy}

The (regularized) internal energy per degree of freedom is
\begin{equation}
    u = \frac{1}{2\beta}\sum_{n\in\mathbb{Z}}\frac{\mu+\Sigma_n}{m\omega_n^2+\mu+\Sigma_n},
\end{equation}
as derived in \cite{Franz13768}. 

\begin{figure}[h]
    \includegraphics[scale=0.8]{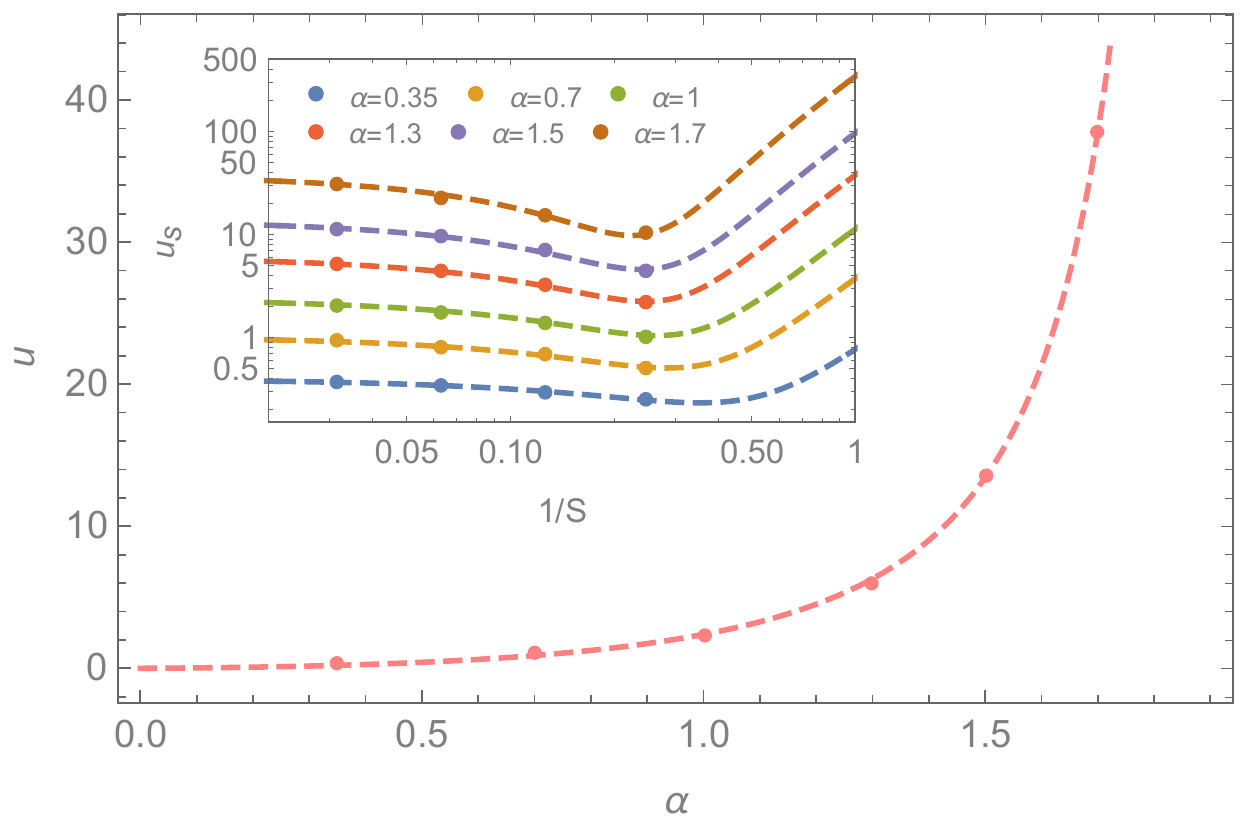}
    \caption{Internal energy $u$ as a function of the density of constraints $\alpha$. The dashed line is a fit of the form $u=A(2-\alpha)^{-\kappa}(1+B(2-\alpha)+C(2-\alpha)^2)$ with $\kappa=2.0$ obtained from the behavior of order parameter $q$. This confirms $u \sim (1-q)^{-1} \sim (2-\alpha)^{-2}$ as discussed in the main text. In the inset one can see, from bottom to top for $\alpha=0.35,0.5,0.7,1,1.3,1.5,1.7$, the extrapolation of the values of $u_S=u+a/S+b/S^2$ as a function of the number of Trotter slices $S$ (in log-log scale).}
    \label{fig:UA}
\end{figure}

We find that $u$ is independent of $\beta$, like $q$, already at $\alpha\gtrsim 1$, but it strongly depends on the number of Trotter slices $S$. Extrapolating the data for $S\to\infty$ we obtain the result in Fig.\ \ref{fig:UA}, which shows a divergence of the energy as $\alpha\to 2$. This is again interpreted in terms of reduced volume and uncertainty principle. In particular, we observe that $u \sim \frac{\hbar^2}{m (2-\alpha)^2}$ with good accuracy for $\alpha\to 2$, in a region where the dependence on $\beta$ is lost. This confirms the result $\kappa \simeq 2$, obtained from the behavior of $q$ (Fig.~3, main text).

\end{document}